\documentclass[sigconf]{acmart}
\AtBeginDocument{%
  \providecommand\BibTeX{{%
    \normalfont B\kern-0.5em{\scshape i\kern-0.25em b}\kern-0.8em\TeX}}}

\setcopyright{acmcopyright}
\copyrightyear{2018}
\acmYear{2018}
\acmDOI{10.1145/1122445.1122456}

\usepackage{graphicx}
\usepackage{amsmath}
\usepackage{color}
\usepackage{array}
\usepackage{multirow}
\usepackage{float}
\usepackage[ruled, linesnumbered, noend]{algorithm2e} 
\setcopyright{iw3c2w3}

\begin{document}

\title{Centralized borrower and lender matching under uncertainty for P2P lending}

\author{Soumajyoti Sarkar}
\authornote{Work done while the author was at Arizona State University}
\email{soumajyoti.sarkar@asu.edu}
\affiliation{%
  \institution{Arizona State University}
  \city{Tempe}
  \country{USA}
}

\renewcommand{\shortauthors}{Sarkar, et al.}

\begin{abstract}
  A clear and well-documented \LaTeX\ document is presented as an
  article formatted for publication by ACM in a conference proceedings
  or journal publication. Based on the ``acmart'' document class, this
  article presents and explains many of the common variations, as well
  as many of the formatting elements an author may use in the
  preparation of the documentation of their work.
\end{abstract}



\copyrightyear{2021}
\acmYear{2021}
\acmConference[WWW '21 Companion]{Companion Proceedings of the Web Conference 2021}{April 19--23, 2021}{Ljubljana, Slovenia}
\acmBooktitle{Companion Proceedings of the Web Conference 2021 (WWW '21 Companion), April 19--23, 2021, Ljubljana, Slovenia}
\acmPrice{}
\acmDOI{10.1145/3442442.3451376}
\acmISBN{978-1-4503-8313-4/21/04}
\begin{abstract}
 Motivated by recent applications of sequential decision making in matching markets, in this paper we attempt at formulating and abstracting market designs for P2P lending. We describe a paradigm to set the stage for how peer to peer investments can be conceived from a matching market perspective, especially when both borrower and lender preferences are respected. We model these specialized markets as an optimization problem and consider different utilities for agents on both sides of the market while  also understanding the impact of equitable allocations to borrowers. We devise a technique based on sequential decision making that allow the lenders to adjust their choices based on the dynamics of uncertainty from competition over time and that also impacts the rewards in return for their investments. Using simulated experiments we show the dynamics of the regret based on the optimal borrower-lender matching and find that the lender regret depends on the initial preferences set by the lenders which could affect their learning over decision making steps.
\end{abstract}

\maketitle

\section{Introduction}
Sequential decision making in two sided markets like consumers and producers has been part of bidding in e-commerce platforms like eBay, eBid for a very long time. Not only that, P2P platforms like Prosper in the past allowed lenders to bid on projects for peer microlending until they switched to posted price mechanism \cite{ceyhan2011dynamics}. However, for most peer microlending platforms like Kiva, LendingClub among others, sequential decision making is either not available or limited in its functionality while investor funding cycles generally have a monopoly on who they fund \footnote{https://www.inc.com/christine-lagorio/sam-altman-yc-monopoly.html}. In this paper, we therefore attempt at abstracting the concept of  two sided markets for peer lending where we consider participants  on either side of the market as agents. We consider a centralized platform for peer lending where both sides have a chance to log their own preferences prior to the start of transactions. This option of both side preferences for peer lending is generally not available either due to the dynamic nature of the loan/project postings or that lenders/investors generally have a budget at any given point in time. In this paper, we consider agents on borrowing side to be a single person or a company raising money while the lender side agents could comprise an individual investor or a venture capital organization. Centralized platforms to tackle these issues could ensure that the transactions between borrowers and lenders are not only based on the money that an investor is willing to put and its preferences but a borrower's willingness to accept the investment (these could be due to issues in lender terms\footnote{https://siliconhillslawyer.com/2019/03/03/standard-term-sheets-problem-yc/} or borrower's assessment of the investor profile). As an added caveat, it also allows for potential bias mitigation that can be implicit in such platforms \cite{sarkar2020mitigating}.

Since we consider the matching for a group of projects to be done prior to the start of the projects and in the same time period as is the situation for most crowdsourcing or for-profit lending platforms, the implicit nature of matching causes competition among agents on both side - the borrowers trying to raise money from the same set of investors while the investors trying to invest in the selected projects. Borrowers and investors are not aware of their preferences that can lead to their goal. To resolve the conflict in this competition and to learn the preferences of agents on other side of the market, we introduce sequential decision making into the matching process which allows agents on the lending side to revise their choices in order to get matched with the least regret and based on their interests over time. This nature of competition in markets for resolving conflicts has been studied recently \cite{liu2020competing} where the agent preferences on one side are concealed from the other and so the sequential decision aspect comes into play for preference revisions over time.  As mentioned, in such P2P platforms, there are mainly two sides to the market: the borrowers who want to borrow money from others for their projects or startups and the lenders  who lend money to borrowers. The traditional rule has been that these sides involve in two sided trading following the Dutch Auction Mechanism \cite{kumar1998internet,wei2017market}. The goal for the borrower is to get its budget completely funded and for the lender to get matched to its most preferred borrower among the set of options given.

For the rest of the paper, we lay the foundations of our ongoing work that demonstrates a way to address competition and fair play in such peer lending platforms with ideas from matching markets\cite{roth1993stable}. The rest of the paper discusses some choices that could be made towards formulating the utilities on both sides, the mechanism for sequential decision making over rounds, and finally the tradeoff between the preference revisions and the utilities for the agents which are also tied in some ways. Throughout the paper, we consider agents are not strategic and therefore their preference submissions are honest.

\section{Framework for Peer Lending}

Our model of lending through a market matching perspective is very close to the Shapley-Shubik model of bilateral trade with indivisible goods \cite{shapley1971assignment} where there is a set of buyers or bidders (the lenders in our case) and a set of sellers selling a unit of good (borrowers in our case) and no lender wants more than one unit of the good.  There is a monetary value that a buyer assigns to the seller's good and this relates to the amount of money that a lender is willing to lend to a borrower posting in our case despite what the borrower project funding requirements (which are generally more than an individual lender can contribute) are. 


In terms of the market design for two sided lending, there are three aspects that control how the transactions are performed in the centralized version of our matching design:

\begin{itemize}
    \item \textbf{Utilities}: Traditionally, most lending platforms have focused on static objectives to maximize the returns based on lender portfolio. However, what is often missing from these static objectives is the concept of utility-driven returns. What it means is that the factor of the lenders on the success of the borrower projects extends beyond just the monetary transaction aspect but the overlap between the interests of the lender, the lender contributions in terms of certain externalities, the egalitarian aspect of market sharing in terms of fairness. So agent utilities (and this is in addition to the agent preferences in our model) also decides how the matching is made and the tradeoff between the preferences and utilities based on returns decide the matching as well. This aspect  in machine learning has been studied in terms of credit scores and advertiser returns recently \cite{liu2018delayed}.
    
    \item \textbf{Matching Market Model}: As one of the key components of our lending, we abstract the resource allocation strategy between the borrowers and sellers in  the form of the traditional matching market model but with added constraints. We assume that the lenders and borrowers do not know each other's preferences throughout the time which is in contrast to most matching markets.
    
    \item \textbf{Sequential decision making}: This notion of interactive matching between two sides of a market has been studied within the framework of unknown preferences amongst the sides to the algorithm \cite{emamjomeh2020complexity}. This popularly known setting called the \textit{bandit} setting is a framework used to learn preferences under uncertainty.
\end{itemize}

Keeping the above in mind, we model the lending platform as a market with 2 sides - the lenders denoted by the set of agents $\mathcal{L}$ = $\{l_1, l_2, \ldots l_N\}$ and the borrowers denoted by the set of agents $\mathcal{B}$ = $\{b_1, b_2, \ldots b_K\}$ and we assume that $K << N$. We now have a two-sided market where the agents on the borrowing side each have their  own funding request proposals and their corresponding requested amount which we denote by $c_b$, where $b \in \mathcal{B}$. Similarly, the lenders each have an overall budget $q_l$, where $l \in \mathcal{L}$. In addition, each set of agents on one side of the market have the opportunity to submit their preferred rankings of the agents on the other side of the market to the platform. These preferences can be conflicting - many lenders might prefer to lend to the same borrower, while multiple borrowers may prefer to tie up with the same lenders having specific portfolio and interests. A key assumption as mentioned is that lenders and borrowers do not know each other's preferences

Additionally, we lay out the following desiderata for our market model that allows us to understand how to design mechanisms for getting the best matches for both the borrowers and the lenders. Each lender $l$ can be matched to at most one borrower while each borrower $b$ can be matched to multiple lenders based on the amount $c_b$ requested. That is, we consider the case of many-to-one matching markets \cite{bodine2011peer}, we treat this design as a simplified version of the ideal many-many matching so as to simplify the demonstration of the dynamics of matching over time steps. Additionally, in the final assignment list of a borrower to multiple lenders, a correct assignment entails that the sum of the donated amounts of the lenders be at least as much as $c_b$ of the borrower. Such mechanisms are currently followed in platforms like GoFundMe or Prosper Full Coverage lending model where the borrower only gets the project funded when the sum of amounts lent, match or exceed the requested amount. Also, we assume that a single agent cannot have ties in its preferences over the agents on the other side of the market.




\subsection{Matching objective}

To decide a matching between $\mathcal{B}$ and $\mathcal{L}$, we introduce the binary decision variable $\mathbf{x}$ := $(x_{bl})_{(b, l)\in \mathcal{B} \times \mathcal{L}}$ such that $x_{bl}$ = 1 if the loan from lender $l$ is assigned and accepted by borrower $b$ and 0 otherwise. Next, the borrower $b$ and the lender $l$ submit their utilities $u_b(l)$ and $u_l(b)$ respectively ($\forall b \in \mathcal{B}, \forall l \in \mathcal{L}$) and these values represent the ordering choice amongst the agents (we discuss the utilities in the next section). The preference orders of the borrowers and the lenders can be captured in the following way: $b \succ_j b' \Longleftrightarrow u_j(b) >  u_j(b')$ and $l \succ_i l' \Longleftrightarrow u_i(l) >  u_i(l')$. Then, $u_{bl}$ = $u_b(l)$ + $u_l(b)$. The total utility of a matching $\mathbf{x} \in \{0, 1\}^{|\mathcal{B}| \times |\mathcal{L}|}$ is given by $\sum_{b \in \mathcal{B}} \sum_{l \in \mathcal{L}} u_{bl} x_{bl}$. In our work, we consider a many-one matching where each borrower is matched to multiple lenders and each lender  matched to a single borrower.  Defining a binary decision variable $\mathbf{w}$ := $(w_{b, l})$ $\in \{0, 1\}^{|\mathcal{B}| \times |\mathcal l{L}|}$, the matching objective in the form of the Gale Shapley constraints can be formulated as \textbf{MQ1}:

\begin{equation} \label{eq:opt}
\begin{array}{ll@{}ll}
\text{maximize } \quad &  \lambda_1\sum_{b\in \mathcal{B}} \sum_{l \in \mathcal{L}}u_l(b) x_{bl}   - \lambda_2 \sum_{b\in \mathcal{B}} \sum_{l \in \mathcal{L}} w_{bl} 
    \end{array}
\end{equation}
\[
\begin{array}{ll@{}ll}
\text{subject \  to} & \sum_{b\in \mathcal{B}} x_{bl} \leq 1  \ \ \ \ (\forall l \in \mathcal{L})\\
& \sum_{l\in \mathcal{L}} x_{bl}q_l \geq c_b \ \ \ \  \ (\forall b \in \mathcal{B}) \\
& c_b x_{bl} + c_b\sum_{b' \succ_l b }x_{b'l} + \sum_{l' \succ_{b} l} q_{l'}x_{bl'} \\ & \geq c_b (1- w_{bl})  (\forall l \in \mathcal{L}, \forall b \in \mathcal{B})\\
& x_{bl} \in \{0, 1\} \ \ \ \ ( \forall b \in \mathcal{B}, \forall l \in \mathcal{L}) \\
& w_{bl} \in \{0, 1\} \ \ \ \ ( \forall b \in \mathcal{B}, \forall l \in \mathcal{L}) 
\end{array}
\]

We denote the set of constraints above as \textbf{C1}. The reader can refer to such formulations of matching \cite{abeledo1996stable} and the Gale-Shapley matching algorithm \cite{roth1989college} for further reading. Briefly these constraints satisfy the following: (1) the lenders can only be matched to one borrower, (2) the number of blocking pairs (denoted by $w_{bl}$) should be minimized in accordance with the original stable matching constraints \cite{roth1993stable}, and the borrower's requested amount must exceed the sum of investments from matched lenders. We leave the proof that discusses the relevance of matching constraints to the above formulation out of this paper but one can check with proof by contrapositive to see why failing to satisfy the constraints would fail to respect the stability constraints. Note above that we optimize for the lender utility in Equation~\ref{eq:opt} but we will come back to this setting when we evaluate our matching objective and which also constitutes the need for our IP formulation instead of the traditional Gale-Shapley agent optimal algorithm.

\subsection{Utility constraints}
The matching objective function with constraints as defined in the previous section depends on the preferences that are set by agents on both sides at the start of each phase. We consider the utilities $\mathbf{u}_b$ and $\mathbf{u}_l$ that denote the vector of values for each agent $l$ or $b$ about its preferences of agents on the other side of the market. And as stated before, the ordering depends on the value that the agents estimate prior to matching. Reasons for lender preferences over borrowers could arise from the return on investment (ROI) which could be calculated in a myriad ways using a lot of other factors \footnote{http://blog.lendingrobot.com/research/calculating-financial-returns-in-peer-lending/}, but for the sake of abstraction, we sample $u_l(b)$ from a uniform distribution.  For the borrower, the main reason to prefer one lender over another is the past reputation of the lender (since network effects can significantly accelerate the funding \cite{horvat2015network}) as well as the interest matches (especially in VC funding, the investor liquidation preferences can play a role in startup preferences). As for the lender case, we sample $u_b(l)$ from a uniform distribution.  These utilities have been calculated before in the form of recommendation systems \cite{choo2014understanding}. So in our case, for a lender $l$, all borrowers are not equal and vice versa which is different from previous assumptions \cite{das2005two}.





\subsection{Sequential decision making}
One important point to note here is that the agents on each side are not aware of the preferences of the agents on the other side, which is why the case for competition arises more prominently. In hindsight, if each lender was aware of its preference position among the borrowers, the lenders would include that in their utility for a more optimal matching. So in order to arrive at a preferred matching faster, we perationalize the matching platform with sequential decision making in the form of multi-armed bandits with a centralized matching platform. This notion of centralized matching markets has been studied before in \cite{liu2020competing}.  From the lender's point of decision making, the unknown probability of being  matched, as well as the probability of the borrower listing getting fully funded constitute the uncertain elements at the start of the matching. However for simplifying demonstration, we assume the uncertainty comes from the absence of knowledge of the borrower preferences (or utilities) to the lenders. The matching procedure happens as follows:  the matching occurs repeatedly for multiple time steps  - in each step, the lender has multiple borrowers to rank, however the rewards from the borrower listing constitutes the uncertain component for the lenders. This is a bandit setting \cite{das2005two} where at each round, the platform provides a pseudo-reward  to the lender based on the borrower it is matched to and allows the lender to revise its preference rankings for the next round. 

\begin{algorithm}[!t]
  \KwIn{$\mathcal{B}$, $\mathcal{L}$,  $u_l$, $u_b$, $T$.}
  \KwOut{Matching $\mathcal{M}_{\mathcal{B}, \mathcal{L}}$}
  
  $cu_l(b)$ $\leftarrow$ $\infty$ \ ($\forall b \in \mathcal{B}, \forall l \in \mathcal{L}$) \\
    $T_{b, l}(0)$ $\leftarrow$ 0 ($\forall b \in \mathcal{B}, \forall l \in \mathcal{L}$) \\

  \For{t = 1, 2, \ldots T}{       
      $\mathcal{M}_b, \mathcal{M}_l$ $\leftarrow$ Matching \textbf{MQ1} using $cu_l$ and $u_b$ ($\forall b \in \mathcal{B}, \forall l \in \mathcal{L}$) \\
    \For{l $\in$ $\mathcal{L}$}{
        $m_l(t)$ $\leftarrow$ $\mathcal{M}_l(l)$  /* matched borrower */ \\
        \If{$m_l(t)$ is not empty}{
            $b_m$ $\leftarrow$ $m_l(t)$ \\
            $r$ $\leftarrow$  $\mathcal{R}(u_{b_m}(l))$  /* lender reward */ \\
            Update $\hat{\mu}_l(b_m)$ using $r$ in Equation~\ref{eq:update_eq} \\
                $T_{b_m, l} (t)$ $\leftarrow$ $T_{b_m, l} (t-1)$ + 1\\ 
        }
            \For{b $\in \mathcal{B}$}{
                 $cu_l(b)$ $\leftarrow$ $\hat{\mu}_{l}(b)$ + $\sqrt{\frac{3 \ \mbox{log} \ t}{2 \ T_{b, l} (t)}}$ \\
            }
      }
   
  }
  return $\mathcal{M}_{b}, \mathcal{M}_l$
  \caption{Matching between borrowers and lenders}
  \label{alg:edge_infer}
\end{algorithm}

In what follows, we explain how the reward distributions for each lender are calculated and which lays the path for the exploration of the arms (here the borrowers) by the lenders at each round. At each time step the platform matches lender $l$ with borrower $m_l$ upon which $l$ is deemed to be able to pull the arm successfully and gets to know the reward $\mathcal{R(.)}$ which is a function of the matched borrower $b_m$ and is sampled from a 1-subgaussian distribution with mean $u_{b_m}(l)$. Following this, the lender $l$ updates their empirical mean for $\hat{\mu}_{l}(m_l)$ through the following equation: 
\begin{equation}\label{eq:update_eq}
  \hat{\mu}_{l}(b) = \frac{1}{1+T_{b, l}(t)} \Big[u_l(b) + \sum_{s=1}^t \mathbf{1} \{m_l(s) == b\} \mathcal{R}(u_b(l)) \Big] 
\end{equation} 

where $T_{b, l}(t)$ = $\sum_{s=1}^t \mathbf{1}\{m_l(s) == b\}$ is the number of times borrower $b$ was matched to lender $l$ till time $t$, and $\mathcal{R}(u_b(l))$ denotes the reward received by the lender $l$ from its matched borrower at time $t$. In Algorithm~\ref{alg:edge_infer}, $\mathcal{M}_l$  denotes the storage structure mapping each borrower to a lender $l$ and similarly and $\mathcal{M}_b$ denotes the set of lenders matched to a borrower which is output at each time step by \textbf{MQ1}.  So it is clear that when posed with many borrowers to choose from and in the absence of any certainty about the rewards that the lender can receive at the end of the project, the borrower can explore or exploit given its history. We utilize the Upper Confidence Bound (UCB) design \cite{lai1985asymptotically} where at each time step $t$ the lenders compute the upper confidence bound for each borrower as follows:

\begin{equation}
cu_{l}(b) = 
\begin{cases}
\infty \ \ \ \ \ \ \ \ \  \ \ \ \ \  \ \ \ \ \ \ \ \ \ \ \ \ \  \ \ \ \ \ \ \ \  ,T_{b, l}(t) = 0 \\
 \hat{\mu}_{l}(b) + \sqrt{\frac{3 \ \mbox{log} \ t}{ 2 \ T_{b, l} (t)}} \ \ \ \ \ \ \ \ ,otherwise \\
\end{cases}
\end{equation}

\begin{figure*}[!h]
\centering
\minipage{0.35\textwidth}
\includegraphics[width=5.5cm, height=3.5cm]{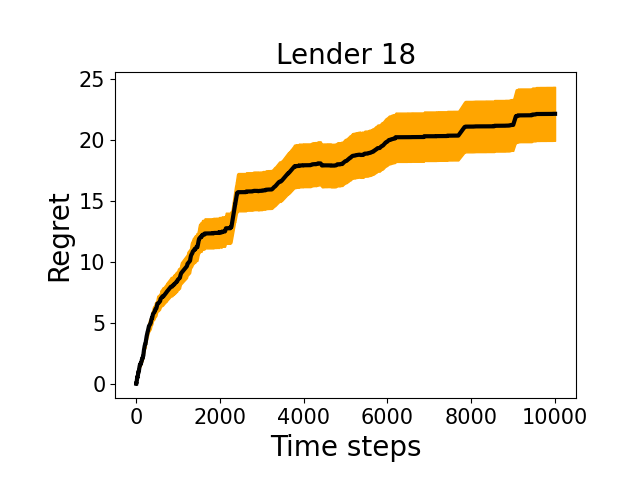}
\endminipage
\minipage{0.3\textwidth}
\includegraphics[width=5.5cm, height=3.5cm]{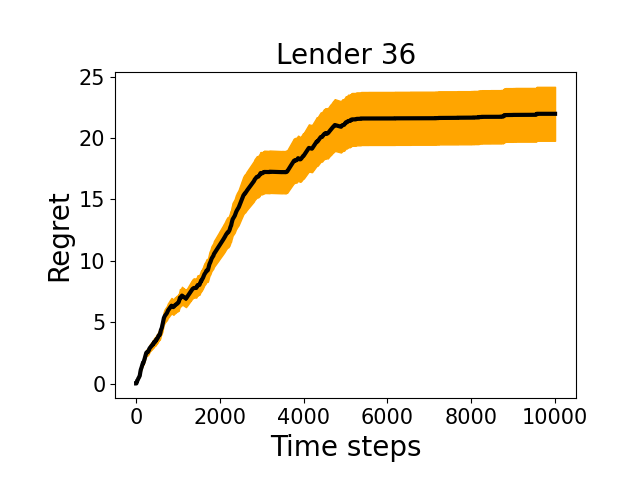}
\endminipage
\hfill
\\
\minipage{0.35\textwidth}
\includegraphics[width=5.5cm, height=3.5cm]{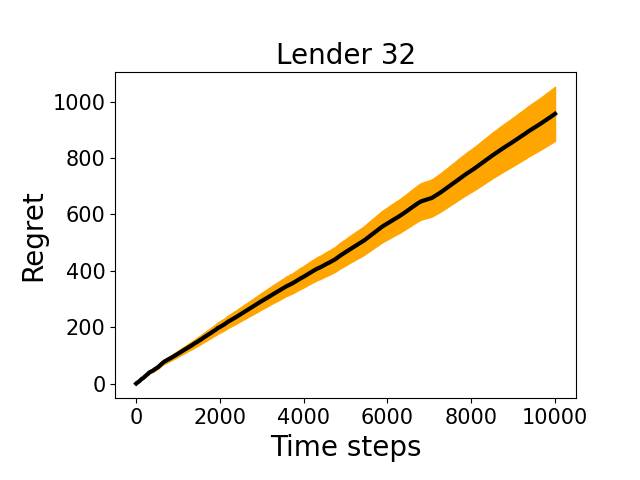}
\endminipage
\minipage{0.3\textwidth}
\includegraphics[width=5.5cm, height=3.5cm]{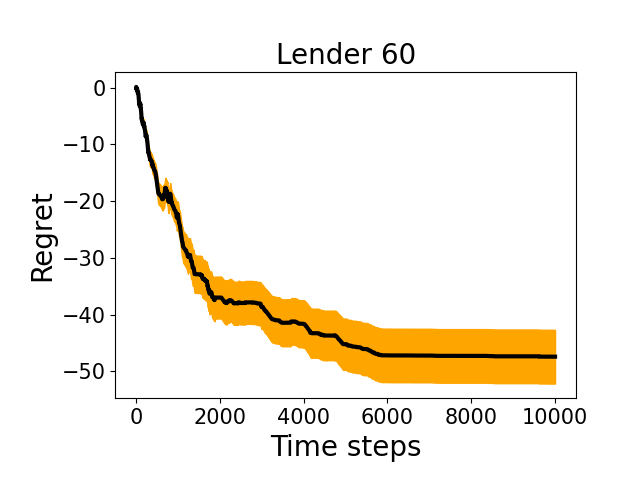}
\endminipage
\hfill
\caption{Cumulative lender regret over 10000 time steps simulated over 50 runs.}
\label{fig:loans_stats}
\end{figure*}

Each lender $l$ ranks the arms $b$ according to $cu_l(b)$ and sends the new utilities to the platform for the next time step while the borrower preferences remain unchanged. \\

\section{Experiments}
We end this paper with a brief example of simulated experiments that lay the foundation for further investigation into what kinds of constraints or better incentives can drive more efficient matchings in future. Our code is fully open sourced and can be accessed \footnote{https://github.com/SOUMAJYOTI/Networked\_bandits}. To simulate the matching, we run the matching algorithm for 40 steps and we consider 20 borrowers and 60 lenders. We sample the borrower and lender utilities $u_b$ and $u_l$ randomly from a uniform distribution (values between 0 and 1). We use Gurobi optimizer to solve the linear program in \textbf{MQ1}.

We randomly sample borrower capacities $c_b$ randomly between values 5 and 40 and for lender budgets we randomly sample values between 1 and 10, but with one constraint, the sum of all borrower capacities must be less than the sum of the lender capacities. And simulate the matching algorithm for 10000 steps over 50 runs. To evaluate the quality of the lender assigned borrowers, we compute the following cumulative regret metric for each lender at each time step $t$:  $t*u_l(b_{opt}) - \sum_{i=1}^t \mathbb{E} \ \mathcal{R}(u_{l}(b_{alg})) $
 where $b_{alg}$ is the matched borrower for lender $l$ at time $t$ returned in Algorithm~\ref{alg:edge_infer} while $b_{opt}$ is computed using the following optimization:  

\begin{equation} \label{eq:opt}
\begin{array}{ll@{}ll}
\text{maximize } \quad & \lambda_1\sum_{b\in \mathcal{B}} \sum_{l \in \mathcal{L}}u_{b, l} x_{bl}   - \lambda_2 \sum_{b\in \mathcal{B}} \sum_{l \in \mathcal{L}} w_{bl} 
    \end{array}
\end{equation}
\[
\begin{array}{ll@{}ll}
\text{subject \  to} &  \textbf{C1} 
\end{array}
\]

Note we optimize the sum of the borrower and lender utilities as in hindsight, the lender would have adjusted its ranking based on borrower preferences had it have access to that information. We list some of the plots of the lender regret over time in Figure~\ref{fig:loans_stats}.we observe two kinds of dynamics from the plots - the lender regret on the top two plots shows that the lender regrets stabilize over time meaning the lenders 18 and 26 have learnt the best preference for them over the period of time. What is more interesting to see is that for some of the lenders like Lender 32, the regret keeps increasing meaning the lenders are unable to learn their preferences over time given their initial utilities and preferences. However, for lender 60 we find that the regret keeps decreasing meaning that it's learnt preference is better than what would be achieved when using the optimal algorithm instead. In such scenarios, our strategy does better than when we optimized for the sum of borrower and lender utilities. One of the reasons behind the lenders being unable to learn might be their initial preference list but also the rewards they obtain over time. Although we modeled the rewards to be proportionate to the borrower-lender utility, this reward function can be modeled on the probability of borrower situation based on the dynamics of all the lenders.

\section{Conclusions and Future work}
We consider a case of a centralized matching platform which requests proposals from borrowers about their preferences over the lenders or the agents through their personal rankings of the lenders. From a lender perspective, this schema thus allows them to get matched without having the information of the actual returns while allowing them certain flexibility to exploit their options.The goal of this paper has been to lay out some ideas in which centralized peer lending platforms can be abstracted from a matching market perspective and how bandits could play a role in mechanism design. These matching markets allow for more privacy as well as ensuring equitable outcomes and in our situation can be achieved by designing proper utility functions of the agents. Similarly in future one could design decision making in which lenders can elicit information about their peer choices as well as networks have been known to aid funding situations \cite{horvat2015network} .

\bibliographystyle{ACM-Reference-Format}
\bibliography{sample-bibliography}

\end{document}